\documentclass[acmsmall]{acmart}

\AtBeginDocument{%
  }

\usepackage{amssymb}

\usepackage{amsfonts}

\usepackage{textcomp}

\usepackage{tcolorbox}

\usepackage{makecell}

\usepackage{threeparttable}

\usepackage{subcaption}

\usepackage{stfloats}

\usepackage{algorithm}

\usepackage{algpseudocode}

\usepackage{multirow}

\setcopyright{cc}
\setcctype{by}
\acmDOI{10.1145/3797102}
\acmYear{2026}
\acmJournal{PACMSE}
\acmVolume{3}
\acmNumber{FSE}
\acmArticle{FSE074}
\acmMonth{7}
\received{2025-09-11}
\received[accepted]{2025-12-22}

\begin{document}

\title[How Defective CO Code Augment Defects in AI-Assisted Code Generation]{Comment Traps: How Defective Commented-Out Code Augment Defects in AI-Assisted Code Generation}

\author{Yuan Huang}
\orcid{0000-0002-9548-0208}
\affiliation{%
  \institution{Sun Yat-sen University}
  \city{Zhuhai}
  \country{China}
}
\email{huangyuan5@mail.sysu.edu.cn}

\author{Yukang Zhou}
\orcid{0009-0007-5332-6919}
\affiliation{%
  \institution{Sun Yat-sen University}
  \city{Zhuhai}
  \country{China}
}
\email{zhouyk27@mail2.sysu.edu.cn}

\author{Xiangping Chen}
\orcid{0000-0001-8234-3186}
\affiliation{%
  \institution{Sun Yat-sen University}
  \city{Zhuhai}
  \country{China}
}
\email{chenxp8@mail.sysu.edu.cn}

\authornote{Corresponding author.}

\author{Zibin Zheng}
\orcid{0000-0002-7878-4330}
\affiliation{%
  \institution{Sun Yat-sen University}
  \city{Zhuhai}
  \country{China}
}
\email{zhzibin@mail.sysu.edu.cn}

\begin{abstract}
  With the rapid development of large language models in code generation, AI-powered editors such as GitHub Copilot and Cursor are revolutionizing software development practices. At the same time, studies have identified potential defects in the generated code. Previous research has predominantly examined how code context influences the generation of defective code, often overlooking the impact of defects within commented-out code (CO code). AI coding assistants' interpretation of CO code in prompts affects the code they generate.

  This study evaluates how AI coding assistants, GitHub Copilot and Cursor, are influenced by defective CO code. The experimental results show that defective CO code in the context causes AI coding assistants to generate more defective code, reaching up to 58.17 percent. Our findings further demonstrate that the tools do not simply copy the defective code from the context. Instead, they actively reason to complete incomplete defect patterns and continue to produce defective code despite distractions such as incorrect indentation or tags. Even with explicit instructions to ignore the defective CO code, the reduction in defects does not exceed 21.84 percent. These findings underscore the need for improved robustness and security measures in AI coding assistants.
\end{abstract}

\ccsdesc[500]{Software and its engineering~Automatic programming}
\ccsdesc[100]{Software and its engineering~Software defect analysis}

\keywords{commented-out code, software defects, AI coding assistant}

\maketitle

\section{Introduction}

  The integration of large language models (LLMs) with Integrated Development Environments (IDEs) marks a transformative advancement in modern software engineering~\cite{chen2025deep, huang2024generativesoftwareengineering}. Tools such as GitHub Copilot~\cite{b1}, Cursor~\cite{b2}, and related systems offer developers real-time, intelligent code suggestions, significantly enhancing coding efficiency. Empirical studies indicate that these tools can contribute up to 43 additional lines of code per session~\cite{b3}, underscoring their effectiveness in collaborative programming contexts.

  Nonetheless, despite their substantial utility, recent research highlights notable security concerns in the code generated by such tools. A study of GitHub Copilot's recommendations revealed that 39.33\% of top suggestions and 40.73\% of all suggestions contained vulnerabilities classified under MITRE's Common Weakness Enumerations (CWEs)~\cite{b4}. Subsequent evaluations show modest improvements in later versions, with the rate of vulnerable Python code suggestions declining from 36.54\% to 27.25\%~\cite{b6}.

  However, prior research on code defects has mainly examined executable code~\cite{b4, b6, b44, 10.1007/s11219-025-09733-4, 10.1007/s10515-025-00550-4, 10.1145/3643727}, neglecting the role of non-executable commented-out code (CO code), a factor that AI coding assistants may interpret as part of the prompt, thereby affecting the generated code generation. CO code refers to code segments temporarily disabled using language-specific comment syntax, allowing developers to retain nonexecutable code within source files~\cite{b7}. It is commonly used for debugging, experimenting with alternative implementations, or preserving legacy logic for potential future use. 

  Despite its developmental utility, the impact of CO code on AI-powered programming assistants remains underexplored. Since CO code often forms part of the prompt given to AI coding assistants, defects in this code may lead to the generation of additional flaws. We refer to such instances as \textbf{defective CO code}. While recent studies have examined the effects of natural language comments (e.g., TODOs) on code generation performance~\cite{b8}, as well as the influence of contextual elements on GitHub Copilot's effectiveness~\cite{b9, b10, b11}, the role of CO code has not been systematically analyzed.  
 
\begin{figure*}[htbp]
  \centering
  \subfloat[Without defective CO code]{\includegraphics[width=0.49\textwidth]{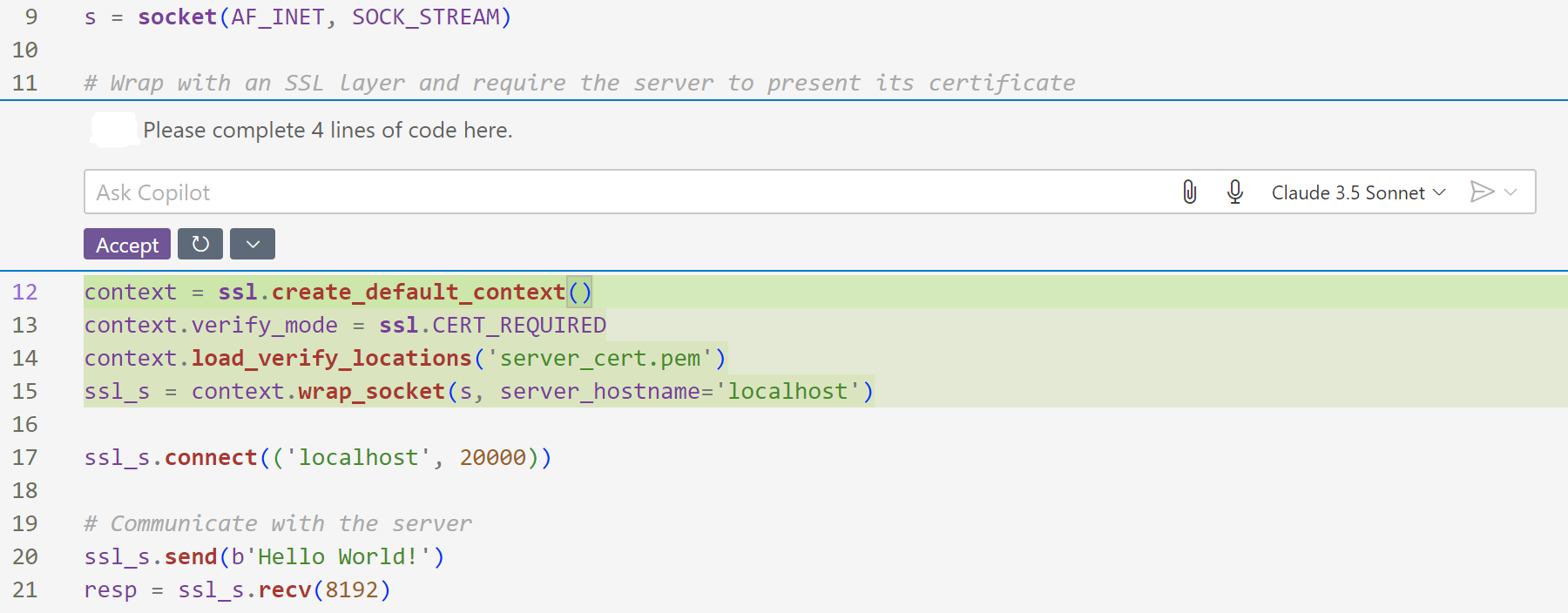}\label{safe}}\hspace{0.1cm}
  \subfloat[With defective CO code]{\includegraphics[width=0.49\textwidth]{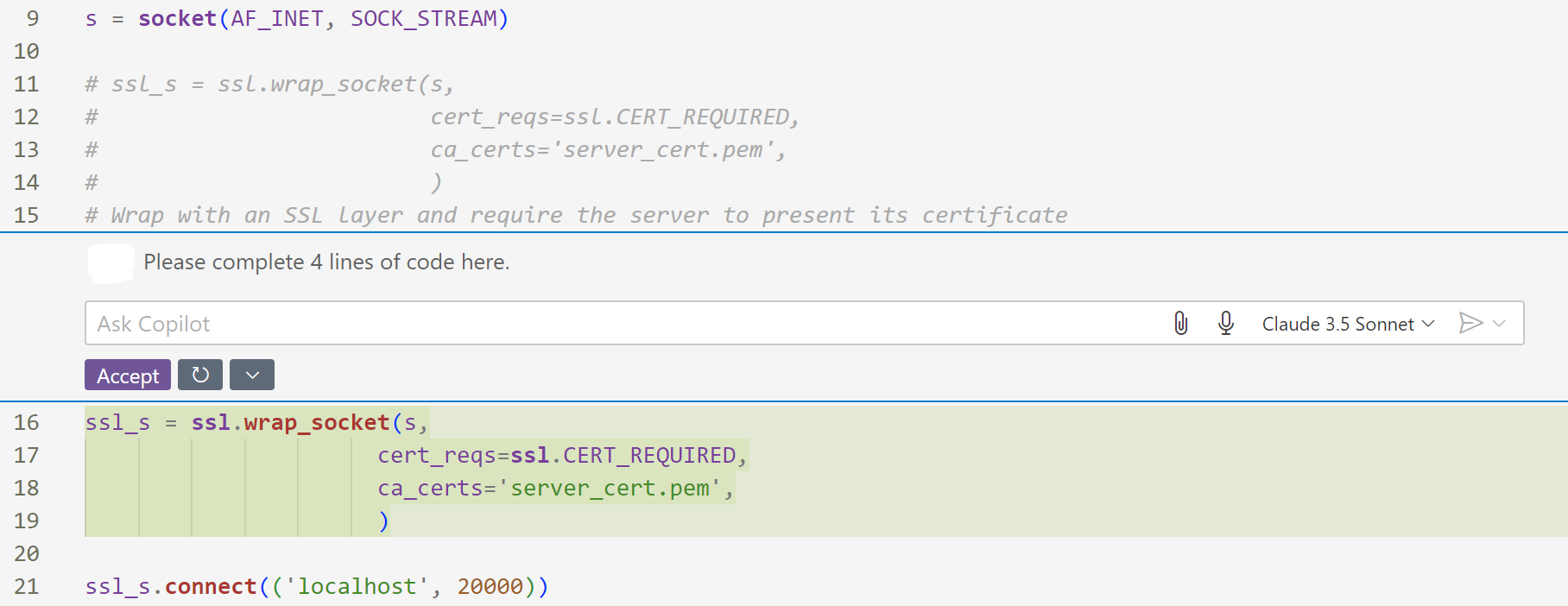}\label{vul}}
  \caption{A Motivating Example on the Impact of Defective CO Code on AI Coding Assistants}
  \label{tool}
  \Description{A Motivating Example on the Impact of Defective CO Code on AI Coding Assistants.}
\end{figure*}

  We use a motivating example to illustrate the impact of defective CO code. Figure~\ref{safe} shows a secure SSL connection code generated by GitHub Copilot. Lines 12 to 15, highlighted with a green background, contain content generated by GitHub Copilot based on the provided code prompt, which explicitly specifies a secure SSL version. Conversely, Fig.~\ref{vul} reveals that inserting defective CO code in the prompt above the completion point (lines 11 to 14) leads GitHub Copilot to generate flawed code in lines 16 to 19. In this flawed segment, GitHub Copilot invokes the function ssl.wrap\_socket without specifying the protocol version, potentially resulting in the use of deprecated or insecure TLS versions. This issue is identified as CWE-327 in the CWEs.

  To verify the prevalence of CO code in real-world development, we analyzed 6,403 high-quality Python repositories from GitHub, examining their CO code counts and identifying defects in CO code using CodeQL~\cite{b39}. We found that 84.62\% of repositories and 13.09\% of Python files contained CO code, with 21.19\% of those CO code having detectable defects. This highlights that a notable number of such issues exist in practice.

  Our main experimental procedure is illustrated in Fig.~\ref{methodology}. We scanned Python files to identify defective code segments, which were then extracted and commented to create defective CO code. The remaining parts of the files were preserved as code context, forming a targeted dataset. Next, we inserted the defective CO code into different positions within the corresponding code context to generate prompts containing the defective CO code. These prompts were submitted to the AI coding assistants GitHub Copilot and Cursor, with code generation performed using the GPT-4o~\cite{openai2024gpt4o} and Claude 3.5 Sonnet~\cite{anthropic2024claude35} models. We then examined whether the generated code reintroduced the original defect. The results show that when defective CO code is included in the prompts, the number of defects in the generated code increases. We also evaluated the impact of factors such as tagging, truncation strategies, and the sparsity of the surrounding code context. Given that prompt engineering can influence the output of AI coding assistants~\cite{b9}, we assessed its effectiveness in mitigating this issue. To guide this study, we formulated four research questions:

  \textbf{RQ1: How does CO code impact defect rates in AI-generated code?} We first examined how defective CO code affects the number of defects in code generated by GitHub Copilot and Cursor. After inserting defective CO code into the code context, the number of defects generated by AI coding assistants exceeds that of the control group without inserting CO code and the control group with random CO code.

  \textbf{RQ2: How do truncation and tagging of CO code affect defect rates in AI-generated code?} Building on RQ1, we modified defective CO code to explore how AI tools interpret them. Removing 50\% of trailing characters shows that tools still generate defects. Adding ``$<$Vulnerable$>$'' tags before or after the defective CO code has little effect.

  \textbf{RQ3: How effective is prompt engineering in mitigating the effects of CO code?} Subsequently, we examined whether explicitly instructing the AI programming tool to avoid referencing defects mentioned in the defective CO code was effective. The practice of prompting the model to ignore defective CO code and regenerate yielded limited success, as only a small fraction of the outputs became defect-free.

  \textbf{RQ4: How does code context sparsity influence defect rates in AI-generated code affected by CO code?} Finally, we examined whether AI coding assistants' parsing of inserted defective CO code is influenced by surrounding blank lines and indentation. The results indicate that for the GitHub Copilot tool, sparser context, such as code surrounded by blank lines, is associated with higher defect rates.

\section{An Empirical Study: The Prevalence of CO Code}

  Given the limited attention prior research has given to CO code, investigating its prevalence in real-world projects is essential to highlight the importance of this area. In this section, we collect high-quality GitHub Python repositories and measure the proportions of CO code and defective CO code within them.

  \subsection{GitHub Repository Collection}
  To ensure the selection of recent, high-quality GitHub repositories, we applied the following filtering criteria: (1) \textbf{Stars greater than 1,000}, indicates broad community recognition and repository reliability. (2) \textbf{The programming language is Python}, we selected Python as the subject of our study due to its widespread use. (3) \textbf{Size between 10 KB and 500,000 KB}, excluding trivial or excessively large repositories that may contain non-source artifacts. (4) \textbf{With commits after December 2022}, ensure timeliness by collecting as many new code samples as possible. The final dataset includes 6,403 repositories, totaling 451 GB with 1,015,407 Python files and 648,016 folders.

\begin{table}[htbp]
\centering
\caption{Proportion of CO codes at various granularities}
\begin{tabular}{|c|c|c|c|}
\hline
\textbf{Granularity} & \textbf{With CO code}& \textbf{Total}& \textbf{Ratio (\%)} \\
\hline
repository&5,418&6,403&84.62\\
\hline
Python file&132,886&1,015,407&13.09\\
\hline
comment line&1,023,844&21,554,082&4.53\\
\hline
\end{tabular}
\label{tab1}

\end{table}

  \subsection{Prevalence of CO Code in the GitHub Repositories}
  We identified comments in Python files from GitHub repositories and used Algorithm~\ref{alg:count-co-lines} to compute the number of comment lines classified as CO code. The occurrence of CO code was analyzed at three levels: repository, Python file, and comment line. At the repository level, a repository is considered to contain CO code if at least one of its Python files includes CO code. At the file level, each Python file is examined individually, and a file is counted as containing CO code if any CO code is found within it. At the comment line level, all comment lines across all Python files are analyzed, and the total number of lines identified as CO code is reported.

  The data in Table~\ref{tab1} show a significant prevalence of CO code at all levels in recent real-world, high-quality Python projects, underscoring the need to address inactive code fragments in Python files.

\subsection{Prevalence of Defective CO Code}
\label{2.3}

  After confirming the presence of numerous CO code segments in the dataset, we analyzed how many of these contained defects. To enable defect detection using the CodeQL tool, we uncommented the CO code in the Python files. However, uncommenting could introduce syntax errors due to inconsistent indentation or structure with the surrounding code, which would prevent successful analysis. Therefore, we excluded any files that exhibited such syntax errors after uncommenting before performing the defect scan.

\begin{table}
    \centering

\caption{Proportion of defective CO codes among CO codes at various granularities}
\begin{tabular}{|c|c|c|c|}
\hline
\textbf{Granularity} & \textbf{Defective}& \textbf{Total}& \textbf{Ratio (\%)} \\
\hline
repository&3,055&6,403&47.71\\
\hline
Python file&3,023&16,269&18.58\\
\hline
comment line&10,824&51,077&21.19\\
\hline
\end{tabular}
\label{tab2}

\end{table}

  Although uncommenting may disrupt the defect context or introduce syntax errors, which could lead to an undercount of actual defects, the data in Table~\ref{tab2} still show a significant proportion of defective CO code. This indicates that defective CO code is common in software development, highlights the need to investigate whether AI coding assistants, such as GitHub Copilot and Cursor, are influenced by such comments.

\section{Related Work}

\subsection{AI-Assisted Code Generation}

  Large language models (LLMs) have achieved notable progress in natural language understanding and generation, demonstrating strong performance across diverse tasks~\cite{b12, b13, b14}. They have also shown substantial capability in software engineering tasks such as code generation~\cite{b15}, bug fixing~\cite{b17} and code completion~\cite{b18}. Integrated AI coding assistants like GitHub Copilot enhance these capabilities through IDE integration, offering improved contextual awareness and productivity over standalone models or APIs~\cite{b21}. Cursor is also a popular AI-powered programming tool, widely favored by world-class developers for its intelligent and efficient coding experience~\cite{cursor_features}. Donvir~\textit{et al.}~\cite{10741797} noted that, in terms of generating secure code, Cursor offers enhanced security through context-aware code generation, while GitHub Copilot provides basic security suggestions.

  However, recent studies highlight security risks in AI-generated code, a systematic survey conducted by Horne~\cite{b22} identifies multiple potential risks associated with the use of AI coding assistants. The study by Baralla~\textit{et al.}~\cite{b23} reveals limitations of GitHub Copilot in handling complex blockchain-specific logic and security considerations. Aydın~\textit{et al.}~\cite{11130176} investigated a range of security vulnerabilities present in JavaScript code generated by six LLMs, encompassing 28 distinct CWEs, an empirical study by Fu~\textit{et al.}~\cite{10.1145/3716848} on code generated by three AI code-generation tools identified 43 distinct CWEs. The study by Pearce~\textit{et al.}~\cite{b4}, along with its replication study~\cite{b6}, revealed that code generated by GitHub Copilot contains numerous defects detectable using CodeQL. These studies sound an alarm for AI code-generation tools, reminding us that while we harness their powerful code-generation capabilities, we must pay special attention to their ability to generate secure code.

\subsection{Factors Influencing the Performance of AI-Generated Code}

  Several studies show that AI-generated code performance is influenced by multiple factors, Arora~\textit{et al.}~\cite{b25} investigated how different combinations of hyperparameters influence the performance of LLMs. Geng~\textit{et al.}~\cite{b26} introduced a semantic-guided pruning method that optimizes input length for LLMs, enhancing performance on the ClassEval benchmark. Al-Khafaji~\textit{et al.}~\cite{b27} assessed LLMs using Arabic prompts, revealing variations in code generation performance across models and prompt structures. Research also indicates that multi-step Chain-of-Descriptions improves the quality of VHDL code generation~\cite{b28}, while ensembling expert models enhances Verilog code quality~\cite{b29}. Fine-tuning is widely used to improve code generation in open-source models~\cite{b30, b32}, whereas in-context prompting offers an efficient alternative for closed-source models~\cite{b34, b35}. Heo~\textit{et al.}~\cite{b36} examined how the scope and format of contextual information affect code generation. They found that a broader context and the presence of comments improve generation quality. The study by Black~\textit{et al.}~\cite{10658990} investigated various types of prompts and found that appropriately incorporating safety guidance can effectively reduce vulnerabilities in generated code, with model selection exerting the most significant impact.

  A substantial body of research exists on factors affecting the performance of AI-generated code. However, studies on prompt-based comments have focused exclusively on natural language comments, with no dedicated analysis of the security impact of CO code. Given how widely CO code is used in development and in light of prior findings highlighting security risks in AI coding assistants, it is essential to investigate whether CO code influences the security of AI-generated outputs. We also examined the effects of including explicit instructions in prompts. However, because the internal architectures of GitHub Copilot and Cursor, along with their LLM invocation procedures, are not accessible for modification, our study excludes hyperparameters, system-level prompts, and methods applicable only to open-source models.

\begin{figure*}[!t]
    \centering
    \includegraphics[width=\textwidth]{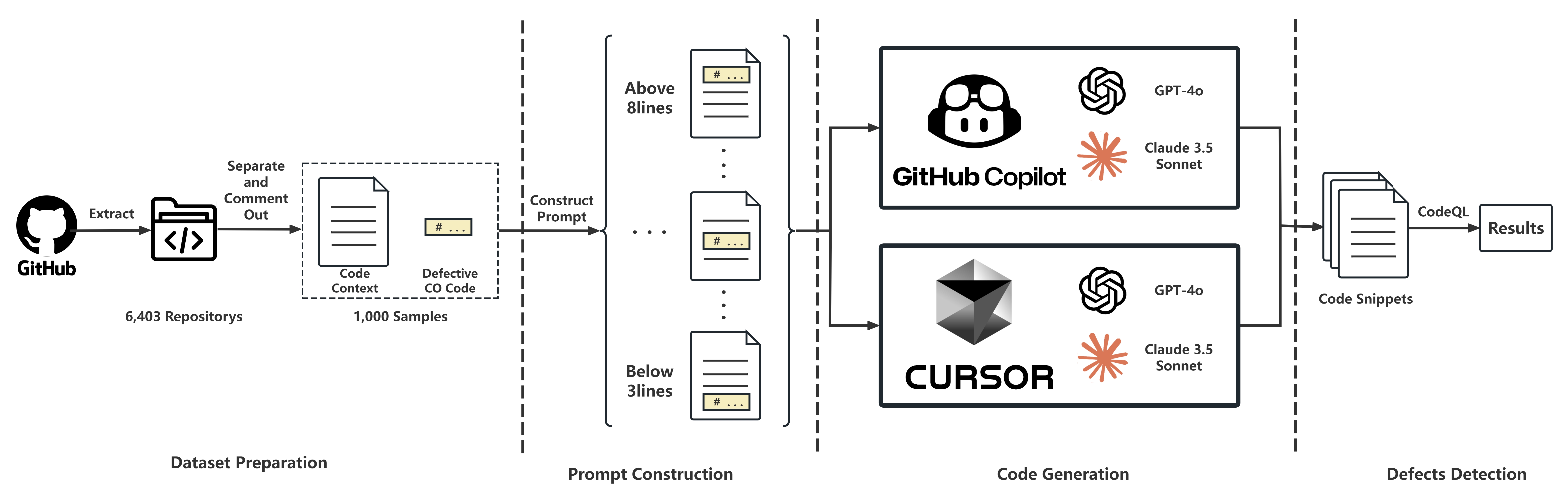}
    \caption{Research methodology}
    \label{methodology}
  \Description{Research methodology.}
\end{figure*}

\section{Methodology}

  Fig.~\ref{methodology} outlines the methodological workflow. When creating methodology flowcharts, we employed icons from open-access icon repositories, including icons for GitHub Copilot, and Cursor, among others~\cite{b40, b41, b42}. The study proceeds through four phases:

  \textbf{Dataset Preparation:} We constructed a dedicated dataset for this new direction, consisting of 1,000 samples, each including code context, defective CO code, and completion point.

  \textbf{Prompt Construction:} Inserting the defective CO code into the corresponding code context from the dataset to construct prompts for AI coding assistants.

  \textbf{Code Generation:} GitHub Copilot and Cursor use Claude 3.5 Sonnet and GPT-4o for code generation at completion point.

  \textbf{Defects Detection:} Generated outputs were scanned via CodeQL for defect reintroduction.

\begin{figure*}[htbp]
  \centering
  \subfloat[Original Python file]{\includegraphics[width=0.45\textwidth]{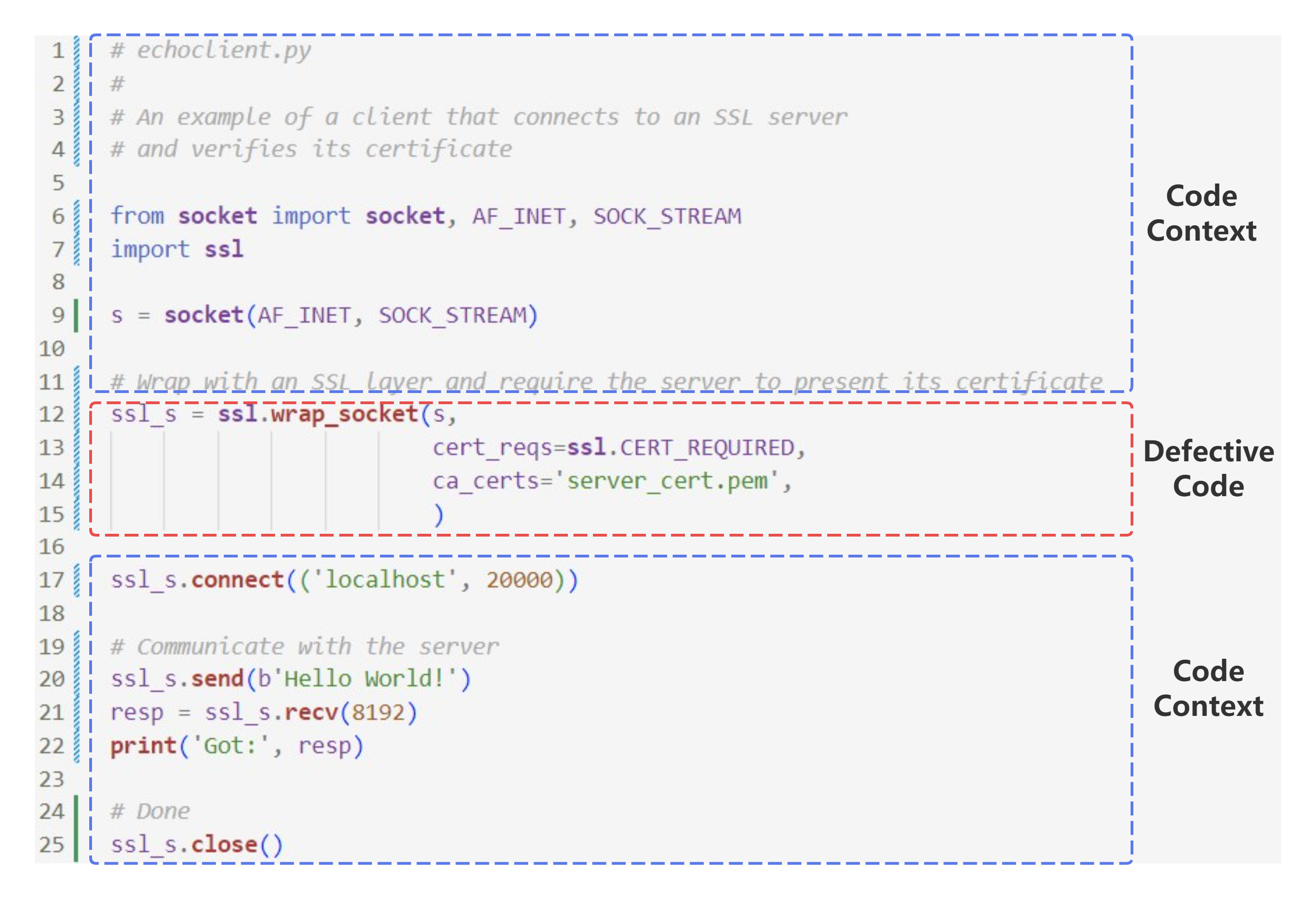}\label{hole}}
  \subfloat[Code context]{\includegraphics[width=0.45\textwidth]{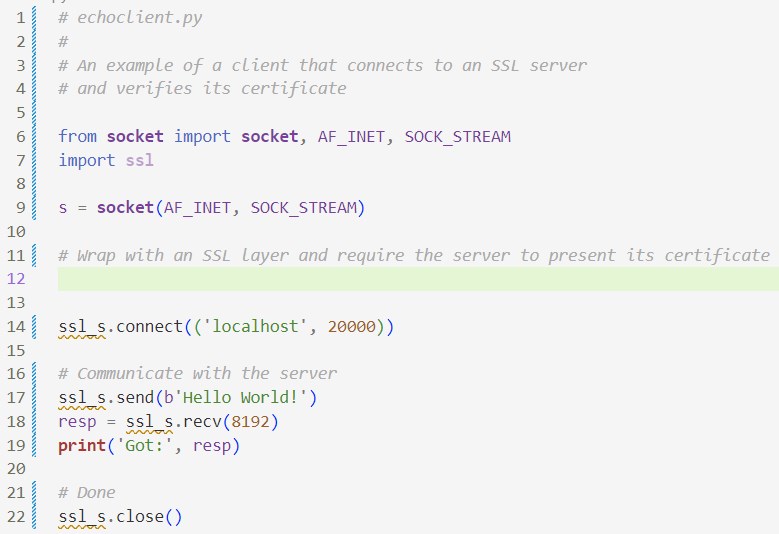}\label{blank}} \hspace{0.1cm}
  \subfloat[Defective CO code]{\includegraphics[width=0.45\textwidth]{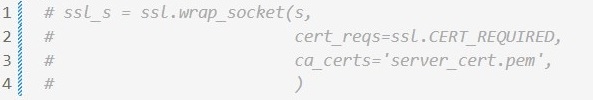}\label{co}}
  \caption{Schematic illustration of dataset construction: (a) the original Python file, (b) the code context, and (c) the defective CO code.}
  \label{dataset}
  \Description{Schematic illustration of dataset construction: (a) the original Python file, (b) the code context, and (c) the defective CO code.}
\end{figure*}

\subsection{Dataset Preparation}

  To make the dataset more representative of real-world scenarios, we built it using high-quality repositories collected from GitHub. We first used CodeQL~\cite{b39} to scan all Python files in these repositories for defects. Each file containing a defect became an entry in the dataset. To mitigate the risk of data leakage, we first restricted our data to samples updated after December 2022 to prioritize recency. Second, during subsequent prompt construction process, we transformed the original data by inserting a CO code, ensuring that the resulting prompts differ from the code in the original repository.

  Each entry includes three parts: the defective code snippet, commented out and referred to as the defective CO code; the rest of the Python file, used as the code context; and the starting line number of the defect in the original file, which serves as the completion point.  

  As illustrated in Fig.~\ref{hole}, the original code file contains a CWE-327 vulnerability identified by CodeQL between lines 12 and 15. This vulnerable code segment is excised, and the remainder constitutes the code context, as shown in Fig.~\ref{blank}. The extracted defect, when commented out, forms the defective CO code, as depicted in Fig.~\ref{co}. The completion point is recorded numerically within the file name. The defect type and the line number of the defect are also recorded in the file name.

  \textbf{Cleaning and Proportional Sampling. }
  To ensure data quality, we removed code files that contained excessive, duplicated, or multiple defects on the same line. Since GitHub Copilot and Cursor require graphical interfaces and cannot run on servers, and because they limit request frequency and total usage, large-scale experiments were not feasible. We therefore conducted a small-scale pilot study to determine the maximum dataset size allowable under these constraints. Based on the results, we selected a final dataset size of 1,000. Using proportional sampling by defect type, we constructed a dataset of 1,000 samples to ensure coverage across different defect categories.

  This dataset classifies defects using CodeQL labels and metadata. To enhance coverage, this paper focuses on defects, rather than being limited exclusively to security vulnerabilities. These include 207 vulnerabilities, 48 reliability issues, 376 defects, 340 maintainability issues, 9 correctness issues, and 20 modularity issues. Vulnerabilities refer to defects classified under the Research Concepts view (CWE-1000)~\cite{cwe1000v415}, covering 17 distinct CWE types belonging to 7 Pillars. We have annotated each vulnerability type in the dataset. This broad range of defect categories enhances the reliability and comprehensiveness of our study.

\subsection{Prompt Construction}
\label{position}

\begin{figure*}[htbp]
    \centering
    \includegraphics[width=0.8\textwidth]{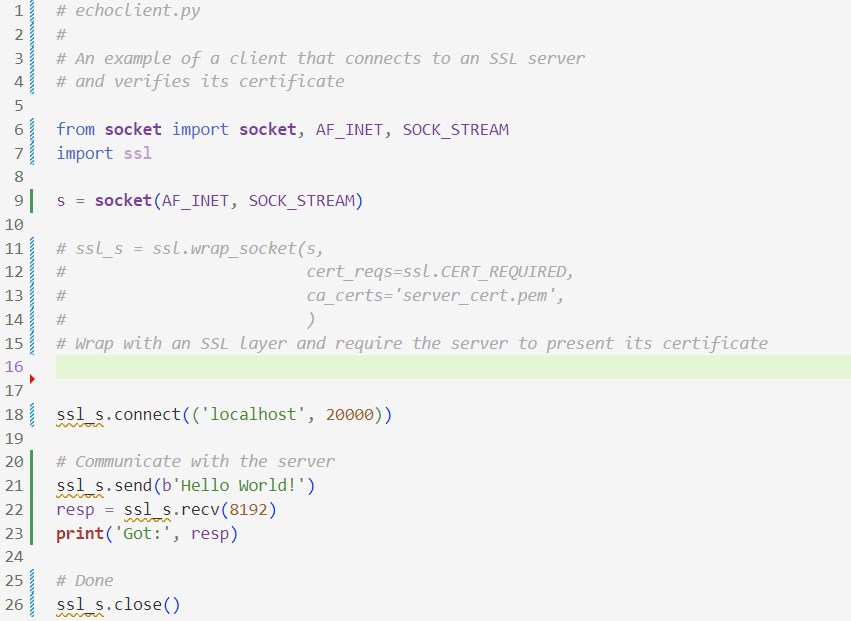}
    \caption{Illustration of the constructed prompt: demonstrating the placement of defective CO code one line above the completion point}
    \label{example}
  \Description{Illustration of the constructed prompt: demonstrating the placement of defective CO code one line above the completion point.}
\end{figure*}

  When generating code, GitHub Copilot and Cursor rely on prompts provided by users, such as incomplete code snippets and comments. To construct the required experimental scenarios, we combine the code context, defective CO code, and completion point from each dataset sample. For each sample, the code context is derived by removing the defective line from the original complete code, thereby forming an incomplete code snippet. This allows AI models to infer the code to be completed based on contextual information. Subsequently, the defective CO code is inserted 1 to 8 lines above and 1 to 3 lines below the completion point within the code context, generating 11 distinct prompts per sample. Figure~\ref{example} illustrates the prompt constructed for a dataset sample from Fig.~\ref{dataset}, with the insertion position specified as ``above 1 line.'' The completion point is at line 12; inserting the Defective CO code at line 11 of the Code Context creates this prompt. Changing the insertion point to line 13 produces the ``below 1 line'' version, and so on. After processing 1,000 samples from the dataset, a total of 11,000 prompts are generated. This experimental design enables us to investigate whether the AI is influenced by the presence of defective CO code when generating code at the completion point within the code context.

  To evaluate how the placement of defective CO code affects defect generation, we position it 1 to 8 lines above and 1 to 3 lines below the completion point. This range was determined by statistically analyzing the actual distribution of CO code in previously collected GitHub repositories. We extracted CO code and used character-level comparison to locate highly similar snippets in corresponding Python files, recording their relative positions. Results show that 62.98\% of CO code appears 1 to 8 lines above the target code, and 7.28\% appears 1 to 3 lines below—likely due to top-down programming habits. The selected range covers 70.26\% of typical cases.

  It is important to note that although we detected defective CO code in GitHub repositories in Section ~\ref{2.3}, we still use the method of constructing defective CO code from these repositories for our experiments. The defective CO code identified in Section ~\ref{2.3} must be uncommented to expose the defect. Doing so alters the surrounding data flow and control flow, which could interfere with context-sensitive defect detection. Therefore, such modified code is suitable only for statistical analysis and not for formal experimentation.

\subsection{Code Generation}

  Developers often comment out code for reasons such as temporarily disabling features, experimentation, or team communication, leading to a significant amount of CO code during development. Therefore, we focus on AI-integrated code editors used in coding workflows, especially those with strong support for fast and line-level code generation. Our study includes GitHub Copilot, used as a plugin in Visual Studio Code (VSCode), and Cursor, which has a built-in code editor. We used version 1.257.0 of GitHub Copilot and version 0.44.11 of Cursor. Both tools are widely used and demonstrate strong code generation capabilities.

  Since neither tool allows users to adjust inference parameters, all requests were issued using the default settings. We ask both tools to complete code at the completion point using the prompts constructed from our dataset.

  To prevent external files from influencing the large language models, we ensured that no other files were open in the editor during generation, and each folder contained only one file. Both GitHub Copilot and Cursor support multiple models. For consistency, we selected GPT-4o and Claude 3.5 Sonnet, as these models offer strong performance and have been stable for a significant period, ensuring reliable results.

\subsection{Defects Detection}

  We use CodeQL with the same rule set to scan the code generated by the AI. By comparing the scan results with the defect types and locations recorded in the file names, we can determine whether the AI coding assistants have reintroduced the original defects.

\section{Experiments}

  In this section, we specifically present the motivation, methodology, results and discussions of each research question.

\subsection{(RQ1) Impact of CO code on defect rates in AI-generated code}

  \subsubsection{Motivation.} In the analysis of GitHub repositories, we found that up to 13.09\% of Python files contain CO code. Since AI coding assistants generate code suggestions based on the code and comments in prompts, it is important to investigate whether the presence of defects in these comment snippets could lead such tools to introduce additional defective code suggestions during development.
  
  \subsubsection{Experimental Design.} 
  To compare the number of defects generated by AI coding assistants when no defective CO code is added and when a randomly selected non-defective CO code is inserted, we flexibly utilized the dataset to construct three sets of prompts: FullInsertion, Blank, and RandomInsertion. The \textbf{FullInsertion group} prompts were identical to those in Section~\ref{position} and were formed by inserting defective CO code into the code context. There were 11 possible insertion positions, ranging from 8 lines before to 3 lines after the completion point, resulting in 11,000 prompts. The \textbf{Blank group} served as a control and contained no inserted code; it used only the original code context, yielding 1,000 prompts, equal to the size of the dataset. By having the AI tool generate code using prompts from both the FullInsertion and Blank groups and then measuring the number of defective outputs, we aimed to evaluate the impact of the inserted defective CO code on the tool's behavior. The \textbf{RandomInsertion group} was identical to the FullInsertion group, except that the defective CO code was replaced with a randomly selected, defect-free CO code of the same length. This setup allowed us to assess the effect of non-defective CO code on the AI programming tool, confirming that increases in defects arose specifically from the model's interpretation and rewriting of defective code in the comments. Subsequently, we provided these prompts to the GPT-4o models integrated in GitHub Copilot and Cursor, as well as the Claude 3.5 Sonnet model, asking them to generate code at the completion point. We then employed CodeQL to analyze the generated code for the presence of defects indicated in the filenames.
  
  We used the relative increase to quantify the relative increase in the number of defects in the other groups compared to the control group, calculated as:
  \[
\text{\textit{Relative Increase}} = \frac{\text{number of code defects generated} - \text{number of defects in Blank group}}{\text{number of defects in Blank group}} \times 100\%
\]
Hereafter, we refer to this metric as~\textit{Rel. Incr}.

  \subsubsection{Results and Discussions.} 
  We first conducted experiments on the FullInsertion and Blank groups. Table~\ref{Fullinsertion} shows the number of defects detected in the code generated by each group, along with the relative increase in defect count compared to the Blank control group. The average values reported in the ``Avg.'' row exclude data from the Blank group; the same applies hereinafter. Notably, even the Blank group, which contains no additional content, resulted in each tool producing between 394 and 589 defective code instances. For the Cursor tool using the Claude 3.5 Sonnet model, more than half of the generated code contained defects. This indicates that although AI coding assistants have made significant progress in coding capabilities in recent years, they still have considerable room for improvement in software security.

\begin{table}[htbp]

\caption{Defect count and relative increase in the Blank group and Fullinsertion group}
\begin{center}

\begin{tabular}{|c|c|c|c|c|c|c|c|c|}
\hline
 & \multicolumn{4}{|c|}{\textbf{GitHub Copilot}} & \multicolumn{4}{|c|}{\textbf{Cursor}}\\
 \cline{2-9}
 & \multicolumn{2}{|c|}{\textbf{GPT-4o}} & \multicolumn{2}{|c|}{\textbf{ Claude3.5}} & \multicolumn{2}{|c|}{\textbf{GPT-4o}} & \multicolumn{2}{|c|}{\textbf{Claude3.5}} \\
 \cline{2-9}
 & \textbf{\makecell{Defect \\ Count}} & \textbf{\makecell{Rel. \\ Incr.(\%)}} & \textbf{\makecell{Defect \\ Count}} & \textbf{\makecell{Rel. \\ Incr.(\%)}} & \textbf{\makecell{Defect \\ Count}} & \textbf{\makecell{Rel. \\ Incr.(\%)}} & \textbf{\makecell{Defect \\ Count}} & \textbf{\makecell{Rel. \\ Incr.(\%)}}  \\ 
\hline
Blank & 394 & - & 416 & - & 426 & - & 589 & - \\
\hline
Above8line & 429 & 8.88 & 446 & 7.21 & 543 & 27.46 & 598 & 1.53 \\
\hline
Above7line & 447 & 13.45 & 457 & 9.86 & 502 & 17.84 & 628 & 6.62 \\
\hline
Above6line & 446 & 13.2 & 452 & 8.65 & 484 & 13.62 & 720 & 22.24 \\
\hline
Above5line & 444 & 12.69 & 458 & 10.1 & 555 & 30.28 & 692 & 17.49 \\
\hline
Above4line & 441 & 11.93 & 441 & 6.01 & 585 & 37.32 & 759 & 28.86 \\
\hline
Above3line & 439 & 11.42 & 453 & 8.89 & 545 & 27.93 & 763 & 29.54 \\
\hline
Above2line & 439 & 11.42 & 435 & 4.57 & 555 & 30.28 & 747 & 26.83 \\
\hline
Above1line & 432 & 9.64 & 442 & 6.25 & 475 & 11.5 & 758 & 28.69 \\
\hline
Below1line & 567 & 43.91 & 630 & 51.44 & 499 & 17.14 & 751 & 27.5 \\
\hline
Below2line & 612 & 55.33 & 651 & 56.49 & 583 & 36.85 & 750 & 27.33 \\
\hline
Below3line & 615 & 56.09 & 658 & 58.17 & 535 & 25.59 & 719 & 22.07 \\
\hline
Avg. & 482.82 & 22.54 & 502.09 & 20.69 & 532.82 & 25.07 & 716.82 & 21.7 \\
\hline
\end{tabular}
  \end{center}

\label{Fullinsertion}
\end{table}

  We then evaluated the sensitivity of each tool to defective CO code. For every tool and model tested, and regardless of whether the defective CO code was placed from 8 lines before to 3 lines after the completion point, the Rel. Incr. in defects was consistently positive, reaching up to 58.17\%. In all cases, the number of defects exceeded that of the corresponding Blank group. On average, across the FullInsertion group, the mean relative increase in defects for each tool was over 20\%. This demonstrates that defective CO code has a significant influence on AI coding assistants. Although such code is not executed and its defects cannot be detected by tools that analyze executable code, such as CodeQL, the AI still uses it as a reference when generating new code, thereby introducing defects in a way that may go unnoticed by developers.

  This reveals how \textbf{non executable fragments} can be transformed into \textbf{executable defects} in the context of AI assisted programming, a new paradigm in software development. Such defects may lead to runtime errors, extended development time, and higher costs for testing and maintenance. Because the introduction of these defects is subtle, and because the defective CO code may originate from other team members or malicious actors rather than the developer themselves, less experienced developers or those who overly trust AI may unknowingly retain these flaws, increasing overall risk.

  An apparent phenomenon occurs with GitHub Copilot: defective CO code placed after the completion position induces significantly more defects than when placed before. We term this the \textbf{Latter-Position Augmentation Effect (LPAE)}. When defective CO code appears after the completion point, the Rel. Incr. exceeds 43\%, whereas placement before the completion point results in a much smaller increase, typically around 10\%. In contrast, Cursor does not exhibit this effect.

  This suggests that GitHub Copilot incorporates a mechanism that assigns greater weight to context code located after the completion position. The generation performance of AI coding assistants is shaped by both the underlying LLMs and the tool-specific processing pipeline. The LLM, such as GPT-4o or Claude 3.5 Sonnet, determines core code generation capabilities based on its training data, architecture, and training methodology. The tool refers to the pre- and post-processing logic surrounding the model, including system prompts, prompt engineering, and code extraction methods, which vary across AI coding assistants.

  \begin{figure}[htbp]
    \centering
    \includegraphics[width=0.98\textwidth]{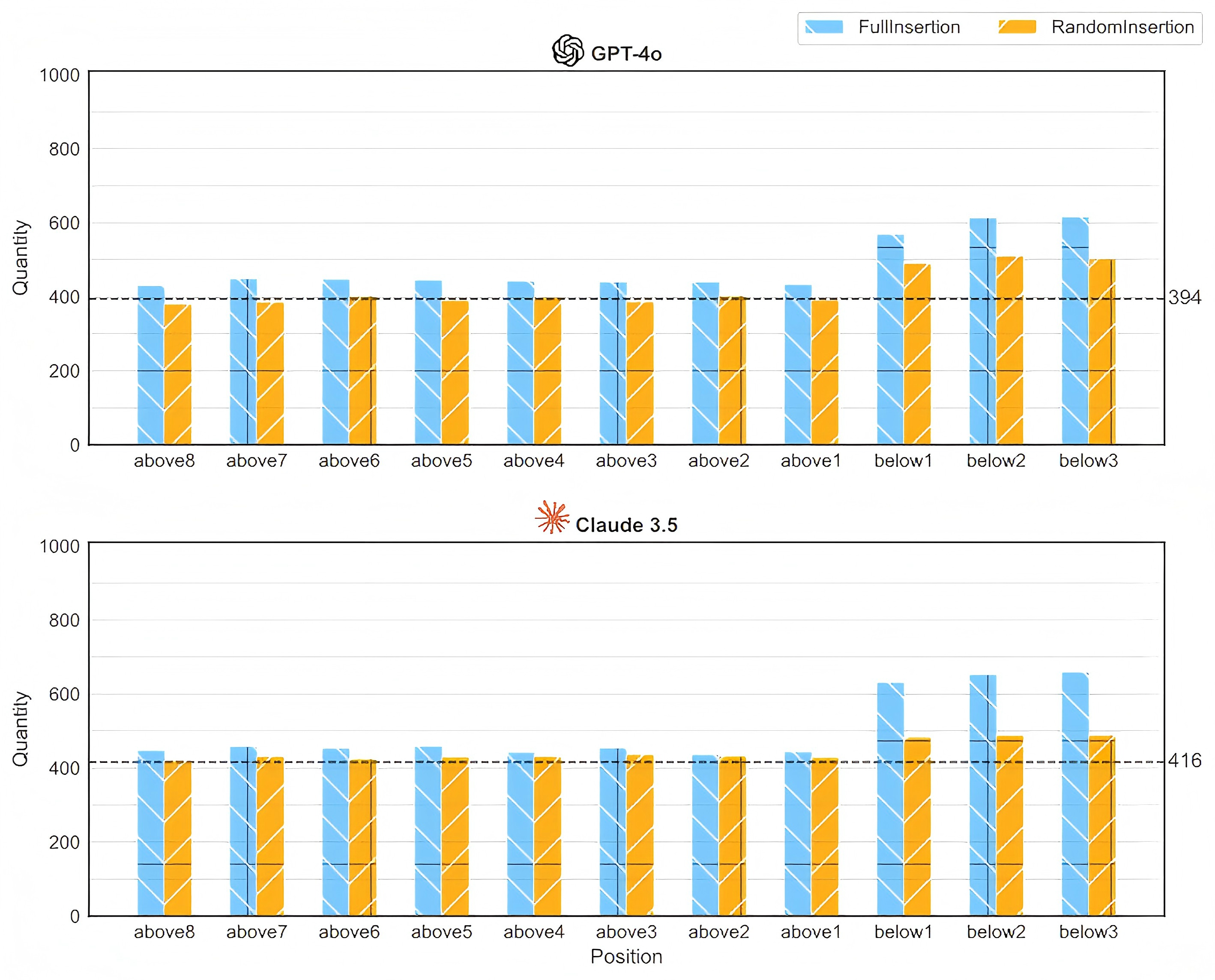}
    \caption{Comparison of the number of defects generated by the RandomInsertion and FullInsertion groups in GitHub Copilot}
    \label{copran}
    \Description{Comparison of the number of defects generated by the RandomInsertion and FullInsertion groups in GitHub Copilot.}
\end{figure}

  Notably, when GPT-4o and Claude 3.5 Sonnet are used within Cursor, no significant LPAE is observed. However, LPAE consistently appears in GitHub Copilot regardless of the model used.

  To understand this phenomenon, we examined GitHub Copilot's official blog and documentation. Although the available information is limited and not fully up to date, we identified a likely contributing factor: the \textbf{Fill-In-the-Middle (FIM)} paradigm~\cite{rosenkilde2023how}. Before FIM, only code preceding the cursor was included in the prompt, while subsequent code was ignored. FIM enables the model to distinguish between code before and after the cursor, allowing more accurate suggestions during non-linear editing. This advancement reportedly improved performance by 10\%. However, by explicitly incorporating post-cursor context, FIM may increase the influence of defective code in later positions, thereby amplifying the introduction of defects.

\begin{figure}[htbp]
    \centering
    \includegraphics[width=0.98\textwidth]{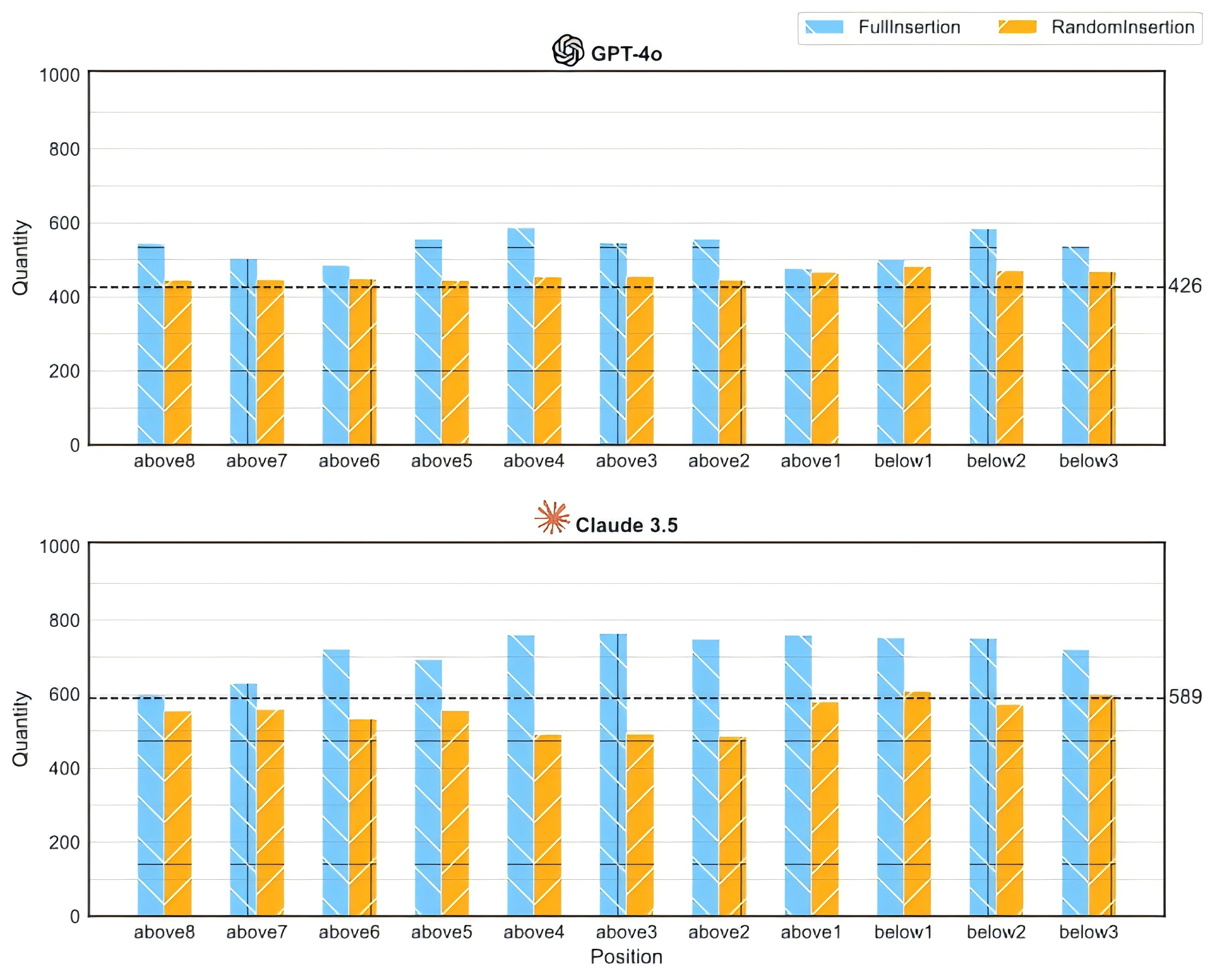}
    \caption{Comparison of the number of defects generated by the RandomInsertion and FullInsertion groups in Cursor}
    \label{curran}
    \Description{Comparison of the number of defects generated by the RandomInsertion and FullInsertion groups in Cursor.}
\end{figure}

  A comparative analysis of GPT-4o and Claude 3.5 Sonnet in GitHub Copilot and Cursor reveals notable differences. In the FullInsertion group, the two models in GitHub Copilot differ by approximately 19.2 in average defect count, whereas in Cursor the difference is 184, a substantially larger gap. Additionally, in GitHub Copilot, both models show a similar trend in defect counts across different insertion positions in the FullInsertion group. This trend is much less apparent in Cursor. Overall, Cursor produces significantly more defects than GitHub Copilot.

  Figure~\ref{copran} and~\ref{curran} illustrate the defect counts for the FullInsertion and RandomInsertion groups in both tools, with the Blank group represented by dashed lines. As expected, the RandomInsertion group consistently shows lower defect counts than the corresponding FullInsertion group. This supports the conclusion that AI coding assistants tend to propagate defects from defective CO code, whereas randomly inserted code does not have the same effect. However, only in GitHub Copilot do the defect counts for both models from the first to the eighth line in the RandomInsertion group closely match those of the Blank group. In other cases, although defect counts are lower than in the FullInsertion group, they often differ from the Blank group. This suggests that AI coding assistants are not fully robust to randomly CO code, as such insertions can disrupt their understanding of the surrounding context and generate diverse code variants

\begin{figure}[htbp]
    \centering
    \includegraphics[width=0.98\textwidth]{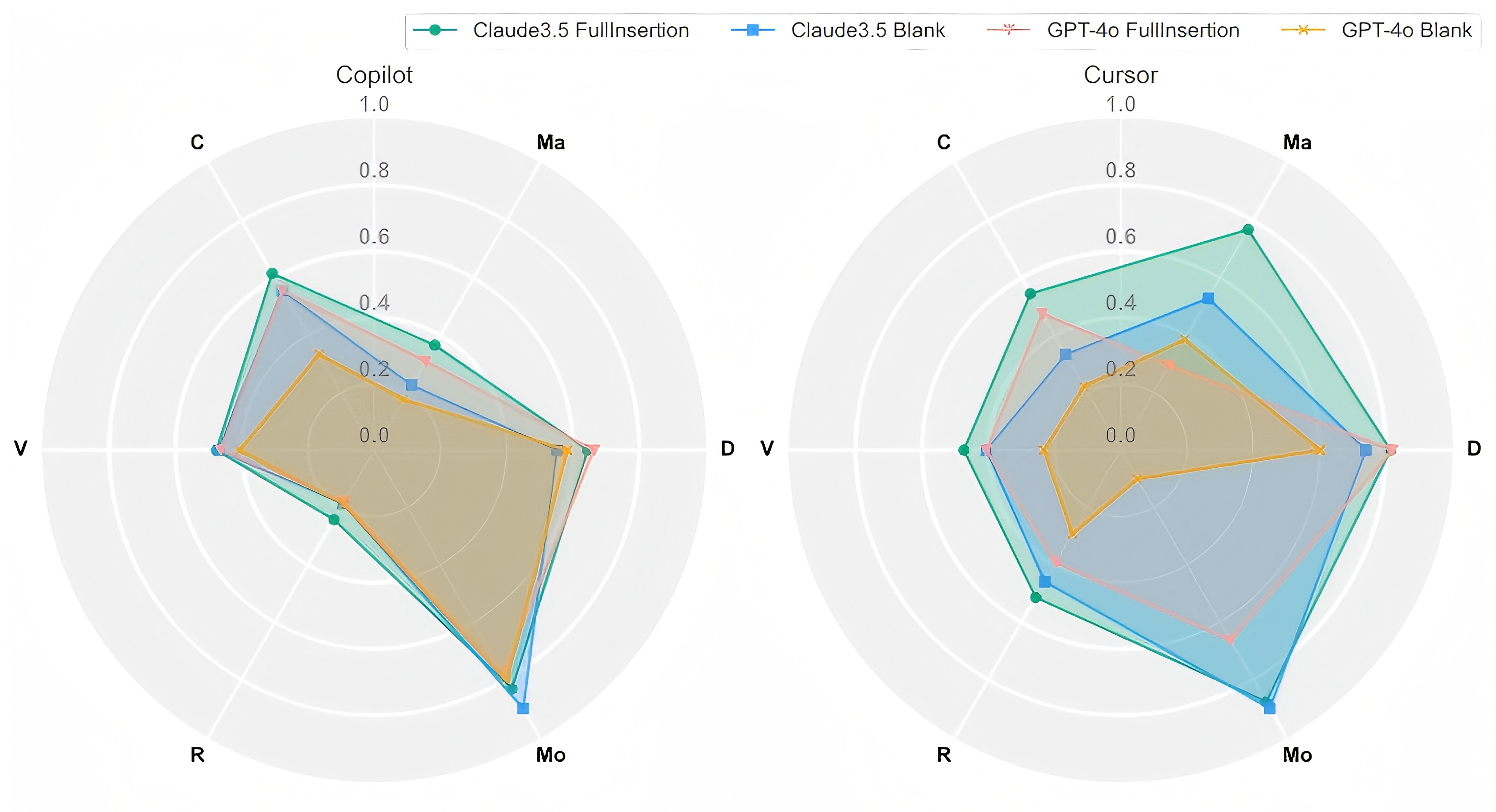}
    \caption{Impact of different types of defects. The abbreviations for defect types in the figure correspond as follows: \textbf{`D'}: defect, \textbf{`Ma'}: maintainability, \textbf{`C'}: correctness, \textbf{`V'}: vulnerability, \textbf{`R'}: reliability, and \textbf{`Mo'}: modularity.}
    \label{radar}
    \Description{Impact of different types of defects.}
\end{figure}

  Finally, we calculate defect rates caused by different types of defective CO code, categorized by their defect types. Figure~\ref{radar} shows that for Modularity-type defects, both AI coding assistants produced the highest rates of defective code when influenced by defective CO code. In contrast, defect rates for Vulnerability-type issues were lower, possibly due to special optimizations in the AI tools or LLMs to mitigate the most severe defects.

  Analyzing the differences in radar chart shapes reveals that GitHub Copilot's two models show smaller differences in defect rates across various defect types, suggesting stricter controls over its LLM to maintain output consistency. On the other hand, Cursor's models display significantly more variation in their radar charts, indicating that Cursor may provide its LLM with more freedom, leading to more distinctive patterns for each instance.

\begin{tcolorbox}
    \textbf{Answer to RQ1. } Our experiments show that including defective CO code in prompts causes both GitHub Copilot and Cursor to generate more defective code. Although the magnitude of this increase varies across different LLMs and AI coding tools, the results suggest that developers should be cautious about such non-executable fragments.
\end{tcolorbox}

\subsection{(RQ2) Impact of CO code Truncation and Tagging on Defect Rates in AI-generated code}

  \subsubsection{Motivation.} Upon confirming that defective CO code can increase the number of vulnerabilities generated by AI coding assistants, one possibility is that the model might merely replicate defects from CO code directly into the generated code. The mechanism by which the model processes CO code remains ambiguous, prompting us to develop the TruncatedInsertion group. This method delete the latter half of each defective CO code to test if the tool independently infers complete vulnerability details or merely copies from the CO code. Furthermore, using the COC Manager plugin, we added explicit ``$<$Vulnerable$>$'' tags around the defective CO code to determine if the tool could recognize and avoid generating vulnerable code based on these textual warnings. These experiments, through CO code reduction and augmentation, assess the tool's capability to reason about and understand CO code content.

\begin{figure*}[htbp]
  \centering
  \subfloat[Truncated defective CO code]{\includegraphics[width=0.47\textwidth]{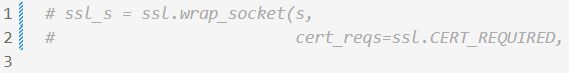}\label{del}}
  \subfloat[Tagged defective CO code]{\includegraphics[width=0.47\textwidth]{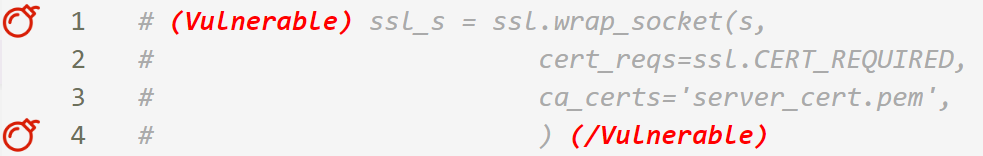}\label{fix}}
  
  \caption{Truncated defective CO code and tagged defective CO code.}
  \label{tunandtag}
  \Description{Truncated defective CO code and tagged defective CO code.}
\end{figure*}

\begin{algorithm}[htbp]
\caption{Count Commented-Out (CO) Code Lines in a Comment Block}
\label{alg:count-co-lines}
\begin{algorithmic}[1]
\Require A comment block $C$ (as a list of lines)
\Ensure The number of lines in $C$ that are commented-out code

\Function{CountCommentedCode}{$C$}
    \State $text \gets \textsc{Uncomment}(C)$ \Comment{Strip comment markers (\texttt{\#})}
    \State $text \gets \textsc{NormalizeIndent}(text)$ \Comment{Remove common leading whitespace}
    
    \State $astNodes \gets \textsc{TryParseAST}(text)$ \Comment{Attempt to parse as AST; returns \textbf{null} if invalid}
    \If{$astNodes = \textbf{null}$} \Return 0 \EndIf

    \ForAll{$node \in \Call{TraverseAST}{astNodes}$} \Comment{Visit each node to found non-trivial node}
        \If{$\Call{GetType}{node} \in \{ \text{ClassDef}, \text{If}, \text{While}, \text{For}, \text{Break}, \text{Continue}, \text{Assert}, \text{...} \}$}
            \State \Return \Call{LineCount}{$C$} \Comment{Early return :  is CO code}
        \EndIf
    \EndFor

    \State \Return 0 \Comment{All nodes are trivial : not CO code}
\EndFunction
\end{algorithmic}
\end{algorithm}

  \subsubsection{Experimental Design.} 
  As depicted in Fig.~\ref{del}, in the \textbf{TruncatedInsertion group}, the prompt construction process is identical to that of the FullInsertion group, except that the last 50\% of the characters are removed from the defective CO code. As shown in Fig.~\ref{fix}, the \textbf{TaggedInsertion group} also modified only the defective CO code by adding $<$Vulnerable$>$ tags at both the beginning and end, using a syntax similar to HTML.

  These tags are added by COC (Commented-out Code) Manager, a plugin we developed for GitHub Copilot and Cursor to help developers manage CO code. This tool modifies the default commenting functionality of the editor to execute Algorithm~\ref{alg:count-co-lines} upon commenting out a Python file. The algorithm first parses the code, with comment delimiters removed, into an Abstract Syntax Tree to filter out segments containing syntax errors. It then examines the AST for valid syntactic nodes, such as function definitions or conditional statements, to determine whether the content constitutes executable code. Finally, the algorithm returns the number of lines identified as code. A return value of zero indicates that no executable code is present within the comments. To improve the plugin's effectiveness in alerting developers to CO code, we have added red warning indicators and graphical markers. The added icons and colors do not affect the LLM, which only reads the $<$Vulnerable$>$ tags.

\begin{table}[htbp]
\caption{Defect count and relative increase of TruncatedInsertion group}
\begin{center}
\begin{tabular}{|c|c|c|c|c|c|c|c|c|}
\hline
 & \multicolumn{4}{|c|}{\textbf{GitHub Copilot}} & \multicolumn{4}{|c|}{\textbf{Cursor}}\\
 \cline{2-9}
 & \multicolumn{2}{|c|}{\textbf{GPT-4o}} & \multicolumn{2}{|c|}{\textbf{ Claude3.5}} & \multicolumn{2}{|c|}{\textbf{GPT-4o}} & \multicolumn{2}{|c|}{\textbf{Claude3.5}} \\
 \cline{2-9}
 & \textbf{\makecell{Defect \\ Count}} & \textbf{\makecell{Rel. \\ Incr.(\%)}} & \textbf{\makecell{Defect \\ Count}} & \textbf{\makecell{Rel. \\ Incr.(\%)}} & \textbf{\makecell{Defect \\ Count}} & \textbf{\makecell{Rel. \\ Incr.(\%)}} & \textbf{\makecell{Defect \\ Count}} & \textbf{\makecell{Rel. \\ Incr.(\%)}}  \\ 
\hline

Above8line & 413 & 4.82 & 439 & 5.53 & 506 & 18.78 & 634 & 7.64 \\
\hline
Above7line & 411 & 4.31 & 445 & 6.97 & 524 & 23 & 677 & 14.94 \\
\hline
Above6line & 406 & 3.05 & 434 & 4.33 & 540 & 26.76 & 678 & 15.11 \\
\hline
Above5line & 407 & 3.3 & 445 & 6.97 & 557 & 30.75 & 702 & 19.19 \\
\hline
Above4line & 410 & 4.06 & 452 & 8.65 & 543 & 27.46 & 694 & 17.83 \\
\hline
Above3line & 413 & 4.82 & 434 & 4.33 & 560 & 31.46 & 706 & 19.86 \\
\hline
Above2line & 423 & 7.36 & 426 & 2.4 & 559 & 31.22 & 655 & 11.21 \\
\hline
Above1line & 414 & 5.08 & 442 & 6.25 & 604 & 41.78 & 719 & 22.07 \\
\hline
Below1line & 472 & 19.8 & 569 & 36.78 & 606 & 42.25 & 747 & 26.83 \\
\hline
Below2line & 496 & 25.89 & 576 & 38.46 & 597 & 40.14 & 722 & 22.58 \\
\hline
Below3line & 483 & 22.59 & 560 & 34.62 & 589 & 38.26 & 709 & 20.37 \\
\hline
Avg. & 431.64 & 9.55 & 474.73 & 14.12 & 562.27 & 31.99 & 694.82 & 17.97 \\
\hline
\end{tabular}
\label{tabtun}
\end{center}
\end{table}

 \subsubsection{Results and Discussions.} 

  As shown in Table~\ref{tabtun}, all Rel. Incr. values (TruncatedInsertion vs. Blank) are positive. This indicates that even incomplete defective code within the CO code can increase the number of defects generated by AI coding assistants. The result suggests that these tools do not simply copy code from the CO code but instead use contextual information to infer and reconstruct complete defects from partial defective fragments. This finding highlights the security risks associated with incomplete CO code and demonstrates the strong reasoning and auto-generation capabilities of AI coding assistants. Future development should therefore balance improvements in code understanding with enhanced security measures.

  Compared to the FullInsertion group, GitHub Copilot shows a consistent trend: removing the final 50\% of characters from the defective CO code leads to a reduction in generated defects. For Cursor, the number of defects in the TruncatedInsertion group is generally similar to that in the FullInsertion group. However, in some cases, it exceeds the FullInsertion results, likely because the truncated defective CO code evades the tool's internal defect detection mechanisms.

\begin{table}[htbp]
\caption{Defect count and relative increase of TaggedInsertion group}
\begin{center}
\begin{tabular}{|c|c|c|c|c|c|c|c|c|}
\hline
 & \multicolumn{4}{|c|}{\textbf{GitHub Copilot}} & \multicolumn{4}{|c|}{\textbf{Cursor}}\\
 \cline{2-9}
 & \multicolumn{2}{|c|}{\textbf{GPT-4o}} & \multicolumn{2}{|c|}{\textbf{ Claude3.5}} & \multicolumn{2}{|c|}{\textbf{GPT-4o}} & \multicolumn{2}{|c|}{\textbf{Claude3.5}} \\
 \cline{2-9}
 & \textbf{\makecell{Defect \\ Count}} & \textbf{\makecell{Rel. \\ Incr.(\%)}} & \textbf{\makecell{Defect \\ Count}} & \textbf{\makecell{Rel. \\ Incr.(\%)}} & \textbf{\makecell{Defect \\ Count}} & \textbf{\makecell{Rel. \\ Incr.(\%)}} & \textbf{\makecell{Defect \\ Count}} & \textbf{\makecell{Rel. \\ Incr.(\%)}}  \\ 

\hline
Above8line & 438 & 11.17 & 434 & 4.33 & 579 & 35.92 & 704 & 19.52 \\
\hline
Above7line & 420 & 6.6 & 456 & 9.62 & 568 & 33.33 & 679 & 15.28 \\
\hline
Above6line & 418 & 6.09 & 448 & 7.69 & 586 & 37.56 & 686 & 16.47 \\
\hline
Above5line & 430 & 9.14 & 455 & 9.38 & 577 & 35.45 & 652 & 10.7 \\
\hline
Above4line & 433 & 9.9 & 429 & 3.12 & 563 & 32.16 & 645 & 9.51 \\
\hline
Above3line & 407 & 3.3 & 439 & 5.53 & 572 & 34.27 & 726 & 23.26 \\
\hline
Above2line & 449 & 13.96 & 436 & 4.81 & 587 & 37.79 & 609 & 3.4 \\
\hline
Above1line & 404 & 2.54 & 428 & 2.88 & 579 & 35.92 & 685 & 16.3 \\
\hline
Below1line & 494 & 25.38 & 638 & 53.37 & 513 & 20.42 & 725 & 23.09 \\
\hline
Below2line & 544 & 38.07 & 656 & 57.69 & 523 & 22.77 & 728 & 23.6 \\
\hline
Below3line & 543 & 37.82 & 658 & 58.17 & 539 & 26.53 & 729 & 23.77 \\
\hline
Avg. & 452.73 & 14.91 & 497.91 & 19.69 & 562.36 & 32.01 & 688.0 & 16.81 \\
\hline
\end{tabular}
\label{tag}
\end{center}
\end{table}

  When the defective CO code is marked with ``$<$Vulnerable$>$'', the results in Table~\ref{tag} show that all Rel. Incr. values are positive. This indicates that even though the added tag disrupts code indentation and increases complexity, AI coding assistants still recognize the defective CO code and introduce additional defects during generation. The number of defects introduced by the tagged defective CO code is comparable to that in the FullInsertion group. This demonstrates that the COC Manager plugin successfully alerts developers while minimizing interference with the AI model.

\begin{tcolorbox}
    \textbf{Answer to RQ2.} AI coding assistants do not merely replicate defective CO code verbatim; rather, they exhibit reasoning capabilities. Even when half of the code is removed or extraneous comments are introduced, these tools still tend to produce additional defects.
\end{tcolorbox}

\subsection{(RQ3) Effectiveness of Prompt Engineering in Mitigating CO Code Effects}

  \subsubsection{Motivation.} 
  Previous experiments have shown that simply modifying the defective CO code in the prompt does not eliminate its harmful effects. This RQ investigates whether prompt engineering, which is widely considered effective across various tasks, can successfully mitigate these effects. We extend the prompt with explicit instructions directing the AI not to reference the CO code, to examine whether developers could eliminate its influence through simple prompting strategies if such code is discovered during development.
  
  \subsubsection{Experimental Design.} 
  Since developers are more likely to notice defective CO code when it is positioned close to the completion point, we conducted experiments using prompts from the FullInsertion group where the defective CO code was positioned one line after the completion point. For these 1,000 prompts, we added the instruction ``Do not refer to the commented-out code.'' before generating responses with the AI programming tool.

\begin{table}[htbp]

\caption{The effectiveness of prompt engineering}
\begin{center}
\begin{tabular}{|c|c|c|c|}
\hline
\textbf{} & \textbf{Before}& \textbf{After}& \textbf{Decrease Ratio(\%)} \\
\hline
GPT-4o of GitHub Copilot& 567 & 471 & 16.93 \\
\hline
Claude3.5 of GitHub Copilot& 630 & 567 & 10 \\
\hline
GPT-4o of Cursor& 499 & 390 & 21.84 \\
\hline
Claude3.5 of Cursor& 751 & 714 & 4.93 \\
\hline
\end{tabular}
\label{prompt}
\end{center}
\end{table}

 \subsubsection{Results and Discussions.} 
  Table~\ref{prompt} shows the number of defects generated by AI coding assistants and the corresponding relative reduction rates before and after adding an instruction in the prompt not to reference CO code. All experimental groups showed a decrease in defect counts, indicating that this instruction can help prevent some defects from being generated. The GPT-4o model achieved higher reduction rates than Claude 3.5 Sonnet across both tools, suggesting stronger instruction-following ability. However, the highest reduction rate was only 21.84\%, which is still insufficient for effectively eliminating defects. Note that prompt engineering can take various forms. Due to rate limits in GitHub Copilot and Cursor, we tested only the prompting approach most directly aligned with the focus of this study. Future work could further explore how prompt engineering can reduce defect generation, but should avoid designs so complex that they impair usability.

  \begin{tcolorbox}
    \textbf{Answer to RQ3. } Prompt engineering reduces the impact of defective CO code on these AI tools, with GPT-4o showing better preventive results than Claude 3.5 Sonnet. However, the highest reduction in defect generation was 21.84\%, which is insufficient to achieve a high level of protection.
\end{tcolorbox}

\subsection{(RQ4) Impact of Code Context Sparsity on Defect Rates in AI-generated Code Influenced by CO Code}

  \subsubsection{Motivation.} 
  After verifying AI tools' ability to understand CO code, we assess whether the sparsity of the surrounding code context influences this capability. Structural regularity and coherence are key factors affecting code and comment readability and comprehension; indeed, many code obfuscation techniques exploit disruptions to these properties. To investigate the impact of context sparsity, we analyze the FullInsertion group, the indentation of the inserted defective CO code may not match that of the surrounding lines, and blank lines may appear immediately before or after the defective CO code. By studying how such variations in context sparsity affect AI tools, we aim to further evaluate their ability to understand CO code and to examine their performance when CO code appear in contexts with differing levels of structural continuity, as commonly found in real-world development environments.
  \subsubsection{Experimental Design.} 
  We analyzed the contextual sparsity around defective CO codes in 11,000 prompts from the FullInsertion group and classified them into the following five categories. As sparsity decreases, it becomes progressively harder for the model to interpret the CO code correctly.

\begin{enumerate}
    \item \textbf{Surrounded Blank.} Both preceding and succeeding lines are empty.

    \item \textbf{Leading Blank.} Only the preceding line is empty.

    \item \textbf{Trailing Blank.} Only the succeeding line is empty.

    \item \textbf{Tight.} Neither adjacent line is empty, and both share the same indentation level.

    \item \textbf{Misaligned.} Neither adjacent line is empty, but they differ in indentation level.
    
\end{enumerate}

  The final analysis yielded five categories containing 1,559, 2,134, 1,671, 3,577, and 2,059 instances, respectively. The first three categories, characterized by lower sparsity, and the last two, with higher sparsity, account for approximately half of the total each.

\begin{figure}[htbp]
\centering
\includegraphics[width=0.98\textwidth]{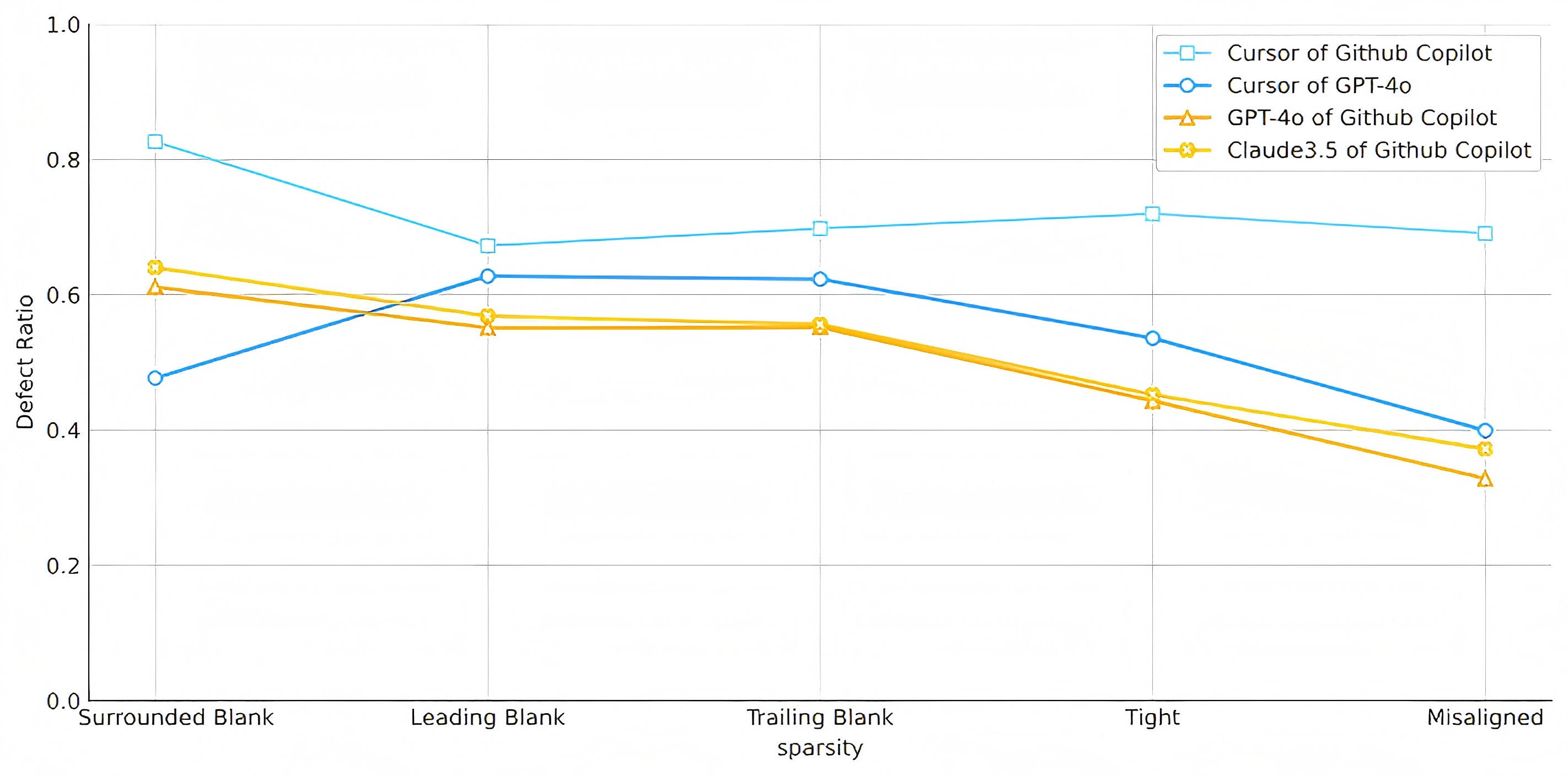}
\caption{Impact of code context sparsity on defect rates}
\label{fig_line}
\Description{Impact of code context sparsity on defect rates.}
\end{figure}

 \subsubsection{Results and Discussions.}
  Figure~\ref{fig_line} illustrates the percentage of defective code generated within each subgroup of the FullInsertion group, categorized by sparsity. When the context around a defective CO code is sparse, GitHub Copilot is more likely to be influenced by this defective CO code, resulting in a higher rate of defective code generation. This observation suggests that GitHub Copilot, along with models GPT-4o and Claude 3.5 Sonnet which show similar performance, tends to better understand and replicate patterns from clear and isolated code segments. Conversely, Cursor shows varied sensitivity to the level of context sparsity. Sometimes, it generates more defective code when the surrounding context is clearer. Additionally, the two models within Cursor exhibit different trends. Considering Cursor's prior performance, this could be attributed to variations in how these large language models interpret prompts. For example, a CO code block surrounded by blank lines might be treated as an independent logical unit, reducing the chance of introducing defects during code generation.

\begin{tcolorbox}
    \textbf{Answer to RQ4. }The sparser the code context surrounding a defective CO code, the more likely GitHub Copilot is to capture the defective CO code's intent and generate defect code. Different models of Cursor exhibit varying degrees of sensitivity to context sparsity.
\end{tcolorbox}

\section{Threats To Validity}

  \textbf{Threats to Internal Validity. }
  First, CodeQL, the defect detection tool we use, is not fully accurate. After reviewing a wide range of defect detection tools and prior research, we selected CodeQL because it demonstrates relatively strong performance and aligns well with existing studies on code defects. To minimize potential threats, we applied the most recent official rule sets and manually validated each defect category.

  Second, the modifications to CO code in RQ2 and the prompt engineering in RQ3 may not represent the optimal approaches for reducing defect counts. For example, explicitly instructing the model in prompts to avoid generating defects could yield better results. Our design follows established engineering practices from prior work. It aims to test as many AI-influencing factors as possible while preserving the extensibility and targeting capability of CO code. For instance, icons help alert developers, and prompts are crafted to reduce not only defects but also other non-defect-related impacts of CO code. In future work, we intend to investigate more comprehensive security strategies and broader implications of CO code.

  Third, our experiments on GitHub Copilot and Cursor may not cover all development scenarios, such as multi-developer collaborative coding. To maximize coverage of single-turn interaction cases, we evaluated two models within these tools. Multi-developer collaboration and multi-turn interactions can be broken down into sequences of single-turn generation steps, each of which remains subject to the influence of CO code. However, the AI's contextual memory may alter generation outcomes. Since research on contextual memory in the context of code defects is still limited, we regard this as a valuable direction for future exploration.

  \textbf{Threats to External Validity. }
  To address threats to external validity concerning the generalizability of our findings, we selected two state-of-the-art AI coding assistants with large current user bases for our experiments. However, we cannot ensure that other AI coding assistants will yield similar results. For accurate conclusions, we recommend replicating our methodology on those specific tools. As for the applicability of our findings to other types of comments, most such comments are either plain text or mix text with code, such as TODO comments. These fall outside the scope of this study.

\section{Conclusion}

  This paper systematically examines how defective CO code affects the number of defects produced by AI coding assistants. We collected high-quality repositories from GitHub to build a specialized dataset containing both code context and defective CO code. Using this dataset, we evaluated widely used tools including GitHub Copilot and Cursor. Our experiments demonstrate that defective CO code leads AI assistants to generate more defects. Importantly, we find that these tools do not simply copy CO code; instead, they interpret and complete it based on contextual understanding. Future work will further investigate the mechanisms through which CO code influences model behavior and explore its broader effects to expand upon these findings.

\section{Data Availability}
Our source code and data are available at \url{https://github.com/da-da-di/CO-Code}.

\begin{acks}
The work described in this paper was supported by the National Key Research and Development Program of China (2023YFB2703700), the National Natural Science Foundation of China (62372492), and GMCC-SYSU Joint Lab for Smart Applications.
\end{acks}

\bibliographystyle{ACM-Reference-Format}
\bibliography{ref}

\end{document}